\documentclass[%
reprint, aps%
 ,twocolumn%
 ,secnumarabic%
,amssymb, amsmath,nobibnotes, aps, 
floatfix]{revtex4-2}

\usepackage{graphicx}

\newcommand\aapr{Astronomy and Astrophysics Reviews}
\newcommand\physrep{Physics Reports}
\newcommand\sovast{Soviet Astron.}
\newcommand\apjl{Astrophys. J.}
\newcommand\apjs{Astrophys. J. Supp.}
\newcommand\jcap{Journal of Cosmology and Astroparticle Physics}
\newcommand\aap{Astron. Astrophys.}
\newcommand\mnras{Monthly Notices of the Royal Astronomical Society}

\usepackage{bm}%
\usepackage[colorlinks=true,linkcolor=blue]{hyperref}%
\expandafter\ifx\csname package@font\endcsname\relax\else
 \expandafter\expandafter
 \expandafter\usepackage
 \expandafter\expandafter
 \expandafter{\csname package@font\endcsname}%
\fi

\begin{document}




\title{Mirror Dark Sector Solution of the Hubble Tension with Time-varying Fine-structure Constant}
			\author{John Zhang$^a$ and Joshua A. Frieman$^{a,b,c}$}
			\email{jfrieman@uchicago.edu}
			\affiliation{$^a$Department of Astronomy \& Astrophysics, The University of Chicago, Chicago IL 60637\\
			$^b$Kavli Institute for Cosmological Physics, The University of Chicago, Chicago IL 60637\\
			$^c$Cosmic Physics Center and Theoretical Physics Division, Fermi National Accelerator Laboratory, Batavia, IL 60510
              }

\date{January 24, 2023}

\newcommand{\LCDM}{\ensuremath{\Lambda\mathrm{CDM}}\xspace}












	\begin{abstract}
We explore a model introduced by Cyr-Racine, Ge, and Knox \cite{cyrracine2021symmetry} that resolves the Hubble tension by invoking a ``mirror world" dark sector with energy density a fixed fraction of the ``ordinary" sector of $\Lambda$CDM. Although it reconciles cosmic microwave background and large-scale structure observations with local measurements of the Hubble constant, $H_0$, the model requires a value of the primordial Helium mass fraction, $Y_p= 0.170 \pm 0.025$, that is discrepant with observations and with the predictions of Big Bang Nucleosynthesis (BBN). We consider a variant of the model with standard Helium mass fraction but with the value of the electromagnetic fine-structure constant, $\alpha$, slightly different during photon decoupling from its present value. If $\alpha$ at that epoch is lower than its current value by $\Delta \alpha \simeq -2\times 10^{-5}$, then we can achieve the same Hubble tension resolution as in \cite{cyrracine2021symmetry} but with consistent Helium abundance. As an example of such time-evolution of $\alpha$, we consider a toy model of an ultra-light scalar field, with mass $m <4\times 10^{-29}$ eV, coupled to electromagnetism, which evolves after photon decoupling at redshift $z \simeq 10^3$, and that with appropriate coupling appears to be consistent with late-time constraints on $\alpha$ variation and the weak equivalence principle. 

	\end{abstract}
			
\maketitle


\section{\label{s:intro}Introduction}
In recent years, local measurements of the current expansion rate of the Universe, the Hubble constant $H_0$, have differed systematically from inferences for $H_0$ from observations of structure in the Universe in the context of the $\Lambda$+cold dark matter ($\Lambda$CDM) model. For example, the most recent analysis from the SH0ES team finds $H_0 = 73.04 \pm 1.04$ km/sec/Mpc \cite{2022ApJ...934L...7R} based on Cepheid variable stars and type Ia supernovae, more than $5\sigma$ discrepant from the value inferred from the final Planck+$\Lambda$CDM cosmic microwave background (CMB) analysis,  $H_0= 67.4 \pm 0.5$ km s$^{-1}$ Mpc$^{-1}$\cite{Planck:2018vyg}. $H_0$ inferences from measurements of large-scale galaxy clustering in $\Lambda$CDM that rely partly on or are independent of the CMB have yielded values generally consistent with the Planck estimate and also in tension with the SHOES result \cite{2019NatAs...3..891V,2021A&ARv..29....9S}. On the other hand, the Carnegie-Chicago local measurements that rely on the Tip of the Red Giant Branch (instead of Cepheids) and Carnegie Supernova Project supernovae found $H_0=69.8 \pm 0.6 \pm 1.6$ km/sec/Mpc \cite{2021ApJ...919...16F,2019ApJ...882...34F}, consistent with both sets of measurements.


Many theories have been proposed to solve the Hubble tension by invoking new ingredients beyond $\Lambda$CDM; these include early dark energy (EDE) models, which invoke ultra-light scalar fields that dominate the energy density just before the epoch of recombination \cite{Smith_2020}, and models such as interacting dark matter or radiation (e.g.,  \cite{Blinov_2020}), decaying dark matter, primordial magnetic fields, etc; for a summary of the tension and various theoretical approaches, see \cite{Di_Valentino_2021,2022PhR...984....1S,2022JHEAp..34...49A,2021APh...13102605D}. However, many of these models fail to adequately match CMB or large-scale structure measurements or both \cite{2021PhRvD.104l3550P, 2021PhRvD.103l3542S}. 

Recently, Cyr-Racine, Ge, and Knox \cite{cyrracine2021symmetry} proposed a different kind of model to resolve the Hubble tension. 
The model appears promising in that it implements a scaling transformation of cosmologically relevant length scales that leaves CMB anisotropy and large-scale structure observables nearly unchanged, thus preserving the remarkable successes of the $\Lambda$CDM model (for earlier related work, see \cite{Zahn:2002rr}). The model features a dark, hidden sector that interacts with ordinary matter only gravitationally and mirrors the ordinary sector by having the same kinds of ingredients (dark sector photons, dark baryons, etc.) and the same interactions within the dark sector as the ordinary sector (see, e.g., \cite{Blinn82,Blinn83,Khlopov1989,Chacko:2005pe,Ciarcelluti:2010zz}). In its simplest incarnation, the energy density of each component in the dark sector is a fixed fraction $(\lambda^2-1)$ of the energy density of the corresponding component in the ordinary sector, so that the total energy density is changed from the $\Lambda$CDM value $\rho_{\Lambda{\rm CDM}}$ to $\rho=\lambda^2 \rho_{\Lambda{\rm CDM}}$. In the limit of thermal equilibrium, CMB and large-scale structure observables are unchanged from their $\Lambda$CDM values to lowest order if the photon scattering rate around the time of photon decoupling is also scaled from $\sigma_T n_e$ to $\lambda'\sigma_T n_e$, with the symmetry condition $\lambda'=\lambda$; here, $\sigma_T$ is the Thomson cross-section, and $n_e$ is the number density of free electrons \footnote{Note that this scaling symmetry of the cosmological perturbation equations in the limit $\lambda'=\lambda$ is not a symmetry in the sense of Noether's theorem and does not imply a corresponding conservation law. It essentially derives from dimensional consistency of the equations of motion. We thank A. Joyce for making this point.}. This can be understood qualitatively by recalling that the redshift of photon decoupling, $z_{\rm dec}$, is defined as the epoch when photons last scatter, i.e., when $(n_e \sigma_T/H) \simeq 1$. Scaling the scattering rate and the expansion rate by the same amount preserves $z_{\rm dec}$ and, more generally, preserves the photon visibility function, a measure of the probability density that a photon last scatters at redshift $z$, given by  $g(z)=(d\tau/dz)\exp(-\tau(z))$, where the derivative of the optical depth $\tau$ is proportional to the scattering rate. 

Cyr-Racine, et al. find that the Hubble tension can be largely resolved, that is, CMB and large-scale structure measurements made consistent with local measurements of $H_0$, if $\lambda \simeq \lambda' \simeq 1.08$. Since, by the Friedmann equation, $H \propto \sqrt{\rho} \propto \lambda$, this scaling accounts for the difference between the low (CMB/large-scale structure) and high (local) $H_0$ values. The model also requires a scaling of the primordial density fluctuation amplitude. 

In \cite{cyrracine2021symmetry}, $\lambda$ is determined by the energy density (or alternatively the relative temperature) of the mirror sector, while $\lambda'$ is determined by a change in the primordial Helium mass fraction, $Y_p$, from its canonical value, since the free electron number density $n_e=X_e n_H = X_e n_B (1-Y_p) =1.1\times 10^{-5} X_e \Omega_b h^2(1+z)^3(1-Y_p)$ cm$^{-3}$, where $X_e \equiv n_e/n_H$ is the ionization fraction, $n_H$ is the number density of Hydrogen nuclei, $n_B$ is the baryon number density, $\Omega_b$ is the baryon density as a fraction of the critical density, and the dimensionless Hubble constant $h=H_0/$100 km/sec/Mpc \cite{1999ApJ...523L...1S,2000ApJS..128..407S}. The symmetry condition $\lambda'=\lambda$ therefore requires a corresponding decrease in $Y_p$ from its $\Lambda$CDM value to $Y_p=0.170 \pm 0.025$ \cite{cyrracine2021symmetry}, which disagrees at $3\sigma$ with the inferred value from observation, $Y_p = 0.2453 \pm 0.0034$ \cite{2021JCAP...03..027A}, so the model is not consistent with all observations. 

In this paper, we hold $Y_p$ to its canonical value and consider an alternative physical mechanism that can scale the photon scattering rate by the requisite amount. In \S \ref{s:sec2} we consider the possibility that the electromagnetic fine structure constant, $\alpha$, and thus the Thomson cross-section, $\sigma_T=(8\pi/3)(\alpha\hbar c/{m_ec^2})^2$, was slightly different around the time of photon decoupling than at the present and show that this can lead to the appropriate enhancement in the photon scattering rate. A change in $\alpha$ also alters the energy levels of atomic Hydrogen and thus the dynamics of Hydrogen recombination and the ionization fraction, $X_e$. As a result, the requisite change in $\alpha$ from its current value, $\Delta \alpha \simeq-2\times 10^{-5}$, turns out to be of opposite sign to and over an order of magnitude smaller than that naively expected from the above scaling of $\sigma_T$.  In \S \ref{s:sec3}, we consider a simple toy model of an ultra-light scalar field coupled to electromagnetism that relaxes the value of $\alpha$ from its primordial to its current value in a manner consistent with  observational constraints. Such dynamical models of time-varying $\alpha$ have a long history in the context of extensions of the Standard Model and in cosmology  (e.g., \cite{1982PhRvD..25.1527B,Webb_2001,Olive_2002,2002PhRvD..66f3507N,Anchordoqui_2003,2003RvMP...75..403U,2011LRR....14....2U,Landau:2008re,Menegoni:2009rg,Martins:2010gu,2010A&A...517A..62L,2012PhRvD..85j7301M,Planck:2014ylh,chluba1,chluba2}).

\section{ \label{s:sec2}Variation of Fine-structure constant and the Photon Mean Free Path}

We consider a model in which the value of the electromagnetic fine-structure constant prior to some time after photon decoupling, $\alpha$, was different from its currently measured value, $\alpha_0\simeq 1/137$. We define the fractional difference $\delta_\alpha$ such that $\alpha=\alpha_0(1+\delta_\alpha)$, and thus the difference in $\alpha$ between early and late times is $\Delta \alpha=\alpha_0 \delta_\alpha$; we assume $\delta_\alpha \ll 1$ throughout and later show that this condition is satisfied in the scaling symmetry limit.
With this assumption, to lowest order the corresponding fractional change in the Thomson cross-section between early and late times is $\delta_{\sigma_T} \simeq 2\delta_\alpha$. Since the current binding energy of atomic Hydrogen is given by $B_0=\alpha^2 m_e c^2/2 = 13.6$ eV, its fractional change at early times is also given by $\delta_B \simeq 2 \delta_\alpha$. This change in $B$ and in associated Hydrogen energy levels propagates to a change in the ionization fraction $X_e$ and thus $n_e$ at fixed temperature. 

\subsection{\label{s:LTE}Decoupling in Thermal Equilibrium}

To get a first estimate of the change in $X_e$ due to a change in $\alpha$, we assume that electrons, protons, H nuclei, and photons remain in thermal equilibrium until photon decoupling, at temperature $T_{\rm dec}$, defined as the epoch when the photon scattering rate, $\lambda_\gamma^{-1}=n_e \sigma_T$, drops below the expansion rate, $H(T)$. In \S \ref{s:NE} we follow the non-equilibrium evolution of $X_e$, but the approximation here of equilibrium followed by decoupling provides some insight into the expected behavior. 

In thermal equilibrium, the ionization fraction is given by the familiar Saha equation (e.g., \cite{Kolb:1990vq,2020moco.book.....D}), 
\begin{equation}
    \frac{X_e^2}{1-X_e}=\frac{1}{n_B} \left(\frac{m_e T}{2\pi}\right)^{3/2} \exp(-B/T) ~~,
    \label{Saha}
\end{equation}
where the baryon density $n_B=\eta n_\gamma=6\times 10^{-10}(\Omega_b h^2/0.022)n_\gamma=2.5\times 10^{-7}(\Omega_b h^2/0.022)(1+z)^3$, and we are using units in which $\hbar=c=k_B=1$. At early times, at temperatures above the binding energy of Hydrogen, $X_e=1$ to excellent approximation, and the Universe is fully ionized; once the temperature drops well below $B$, $X_e$ plummets to a value $X_e \ll 1$. By the time of decoupling, when  $n_e \sigma_T=H(T_{\rm dec})$, which corresponds to a temperature of $T_{\rm dec} \simeq 0.26$ eV and redshift $z_{\rm dec} \simeq 1100$ in the standard model, $X_e(T_{\rm dec}) \sim 0.01$ (e.g., \cite{2020moco.book.....D}). The equilibrium evolution of $X_e$ is shown by the blue curve in the top panel of Fig. 1 for the Planck-$\Lambda$CDM value of $\Omega_b h^2 = 0.022$.

In the evolving-$\alpha$ model, prior to and including the decoupling epoch the binding energy of Hydrogen is given by $B=B_0 + \Delta B \simeq B_0(1+2\delta_\alpha)$.
Defining $X_{e,0}$ to be the canonical $\Lambda$CDM value of the ionization fraction, the Saha equation becomes
\begin{equation}
    \frac{X_e^2}{1-X_e}=\frac{X_{e,0}^2}{1-X_{e,0}}\exp(-\Delta B/T)~.
\end{equation}
Defining the perturbed ionization fraction by $X_e=X_{e,0}(1+\delta_X)$ and expanding to linear order in $\delta_X$, we find 
\begin{equation}
    \delta_X 
    \simeq \left(\frac{1-X_{e,0}}{X_{e,0}-2}\right) \left[\exp \left(\frac{2\delta_\alpha B_0}{T}\right)-1\right]~.
    \label{dXSaha}
\end{equation}
Finally, around the time of photon decoupling, we have $X_e \ll 1$, and the scaling of the photon inverse mean free path due to change in $\alpha$ is given to leading order by 
\begin{eqnarray}
    \lambda'(T_{\rm dec}) && = \frac{(n_e \sigma_T)_{\rm dec}}{(n_e \sigma_T)_{{\rm dec},0}} = (1+\delta_{\sigma_T}+\delta_X)  \nonumber \\
&&    \simeq \left[1+2\delta_\alpha -\frac{1}{2}
    \left[\exp \left(\frac{2\delta_\alpha B_0}{T_{\rm dec}}\right)-1\right]\right]
    ~.
    \label{lambdadec}
\end{eqnarray}

In the limit that the CMB anisotropy is imprinted instantaneously at the epoch of 
decoupling, what we care about is the value of $\lambda'$ at that epoch, $\lambda'(T_{\rm dec})$. From Eqn.(4), since the scaling of $H$ keeps the decoupling temperature $T_{\rm dec}$ at its canonical $\Lambda$CDM value, we find  $\lambda'(T_{\rm dec})=1.08$ (the required value from  \cite{cyrracine2021symmetry}) for $\delta_\alpha =-1.6\times 10^{-3}$, or $\Delta \alpha = -1.2\times 10^{-5}$. The required variation in $\alpha$ is about 50 times smaller than one would naively expect, due to the exponential term in Eqn. (\ref{dXSaha}). The blue curve in the bottom panel of Fig. 1 shows the evolution of $\delta_X$ from Eqn.(\ref{dXSaha}) for this value of $\delta_\alpha$. 
This is our first rough estimate of the needed variation in $\alpha$ for this model.

Note that, strictly speaking, in the model of \cite{cyrracine2021symmetry} we should have $\lambda'=\lambda$ at {\it all} times in order to preserve the scaling symmetry that keeps the CMB and large-scale structure predictions of $\Lambda$CDM intact. In principle this could be arranged through an appropriately time- or temperature-varying $\delta_\alpha$ that leaves the RHS of Eqn. (\ref{lambdadec}) approximately temperature-independent. 

A simpler dynamical model (see \S \ref{s:sec3})  instead assumes that $\delta_\alpha$ has a fixed, non-zero value in the early universe until at least the epoch of photon decoupling and then relaxes to zero sometime after that. In this case, from Eqn.(\ref{lambdadec}) $\lambda'$ is not constant in time prior to decoupling, that is, the symmetry condition does not hold at all times. We can estimate how much this simpler model of $\alpha$-variation breaks the scaling relation $\lambda'=\lambda$ during the epoch when CMB anisotropies are imprinted. The visibility function has most of its support over the range $z_{\rm dec}=1100 \pm 100$ \cite{2020moco.book.....D}, which corresponds to a temperature range $T_{\rm dec}=0.26 \pm 0.023$ eV. Over this temperature range, from Eqn. (\ref{lambdadec}), for constant $\delta_\alpha$, $\lambda'$ varies by roughly $\pm 0.5\%$, which appears to fall within the parameter uncertainty of \cite{cyrracine2021symmetry}.
We therefore expect a model in which $\delta_\alpha=$ constant prior to and during the epoch of photon decoupling to satisfy the scaling symmetry to the needed level, given current observational uncertainties.

\begin{figure*}[t!]
\includegraphics[width=0.9\textwidth]{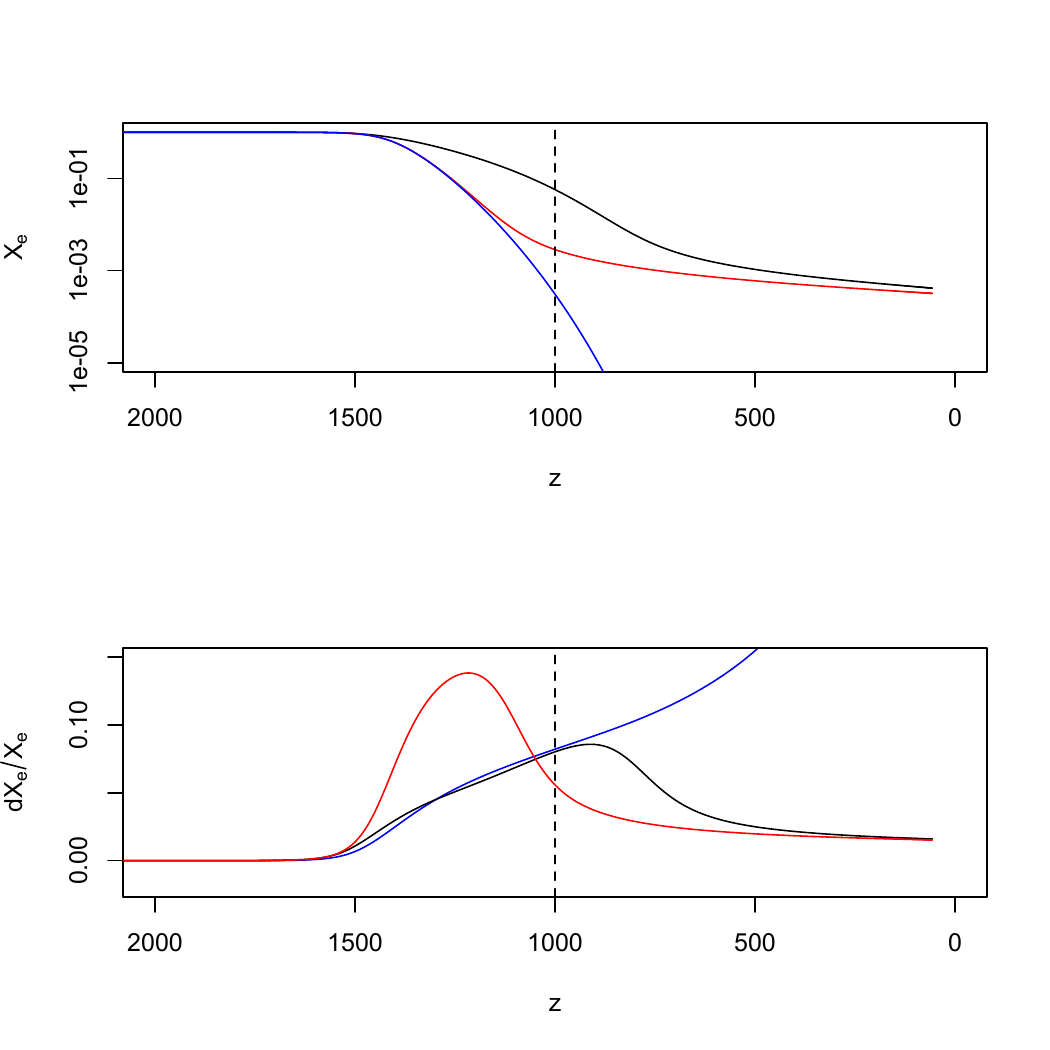}
\caption{\label{fig:1} {\it Top:} Ionization fraction $X_e$ vs. redshift in the standard $\Lambda$CDM model. Blue curve: equilibrium (Saha) solution; red curve: non-equilibrium solution with $C=1$ (see text); black curve: non-equilibrium solution with $C$ factor. {\it Bottom:} Fractional change in ionization fraction, $\delta_X$, vs redshift, due to shift in $\alpha$. Blue curve: equilibrium solution from Eqn.(\ref{dXSaha}) for $\delta_\alpha=-1.6\times 10^{-3}$; red curve: non-equilibrium evolution from differencing solutions of Eqn.(\ref{Xnoneq}) with $C=1$ and $\delta_\alpha=-3\times 10^{-3}$; black curve: same, but including $C$-factor. }
\end{figure*}

\subsection{\label{s:NE}Non-equilibrium Evolution}


The assumption of thermal equilibrium breaks down during the epoch of Hydrogen recombination and photon decoupling. A more accurate estimate of $\lambda'(T_{\rm dec})$ requires solution of the non-equilibrium evolution of the ionization fraction $X_e$. For purposes of illustration, we follow the standard approach based upon the effective three-level atom model \cite{peebles,zeldovich}; approximating the matter and radiation temperatures as equivalent, the evolution is given by 
\begin{equation}
    \frac{dX_e}{dt}= C\left[(1-X_e)\beta - X_e^2 n_B \alpha^{(2)}\right]~,
    \label{Xnoneq}
\end{equation}
where $n_B$ is the baryon density, and following \cite{2020moco.book.....D} we can write the recombination and photoionization rates as 
\begin{equation}
    \alpha^{(2)}=9.78 \frac{\alpha^2}{m_e^2}\left(\frac{B}{T}\right)^{1/2} \ln \left(\frac{B}{T}\right)~,
\end{equation}
and 
\begin{equation}
    \beta = \alpha^{(2)}\left(\frac{m_eT}{2\pi}\right)^{3/2}\exp\left(-\frac{B}{T}\right)~.
\end{equation}
The Peebles $C-$factor can be expressed as (e.g., \cite{2020moco.book.....D}) 
\begin{equation}
    C=\frac{\Lambda_\alpha + \Lambda_{2\gamma}}{\Lambda_\alpha + \Lambda_{2\gamma}+\beta^{(2)}}~,
    \label{Cfactor}
\end{equation}
where the two-photon decay rate is $\Lambda_{2\gamma}=8.227$ s$^{-1}$ and scales as $\alpha^8$ \cite{chluba1}, the Lyman alpha production term is $\beta^{(2)}=\beta \exp(3B/4T)$, and the rate of escape of Lyman-$\alpha$ photons is given by
\begin{equation}
    \Lambda_\alpha = \frac{H(3B)^3}{n_B(1-X_e)(8\pi)^2}~.
\end{equation}

While more realistic and accurate recombination models involving multi-level atoms have been developed and will be invoked below \cite{1999ApJ...523L...1S,2000ApJS..128..407S,2011MNRAS.412..748C,2011PhRvD..83d3513A}, the standard approach again provides some insight into the result.

The black curve in the top panel of Fig. 1 shows a numerical solution of Eqn.(\ref{Xnoneq}) with standard values of parameters. For comparison, the red curve shows the solution with the Peebles $C-$factor of Eqn.(\ref{Cfactor}) set to unity; deviation from the equilibrium solution of Eqn.(\ref{Saha}) becomes apparent around the redshift of decoupling. Since $C<1$ and depends on temperature, its inclusion leads to a delay in the epoch of recombination and thus photon decoupling.

We now consider how a change in $\alpha$ impacts recombination. We set $\alpha \rightarrow \alpha_0(1+\delta_\alpha)$ in Eqn.(\ref{Xnoneq}) and plot the resulting fractional change in $X_e$ in the bottom panel of Fig. 1. The black curve shows $\delta_X$ for $\delta_\alpha=-3\times 10^{-3}$; the red curve shows $\delta_X$ for the same value of $\delta_\alpha$ but setting $C=1$ for comparison. This value of $\delta_\alpha$ yields a peak value of $\delta_X =0.086$ at redshift $z \simeq 910$; in combination with the much smaller variation in $\sigma_T$ this yields a value of $\lambda'(T_{\rm dec})=1.08$, as desired for the scaling solution of the Hubble tension. For $z > 900$, the evolution of $\delta_X$ for this value of $\delta_\alpha$ (black curve) in the non-equilibrium model traces quite well the evolution of $\delta_X$ in the Saha approximation (blue curve) for the smaller value of $\delta_\alpha$ inferred in the previous subsection. 

While the three-level atom model should provide a more accurate estimate than the equilibrium approach of the Saha equation, this model is itself an approximation to a full multi-level approach to recombination \cite{2000ApJS..128..407S,2011MNRAS.412..748C,2011PhRvD..83d3513A}. In fact, the impact of $\delta_\alpha$ in a full multi-level calculation has been considered by a number of authors (\cite{chluba1,chluba2} and references therein). In particular, comparing our results (black curve in Fig. 1) to Fig. 3 of \cite{chluba1} for the same value of $\delta_\alpha$, we find that the three-level calculation agrees qualitatively with their results but appears to {\it underestimate} the peak value of $\delta_X$ by about 7.5\%. With this recalibration, we find that the desired value of $\delta_\alpha=-2.7\times 10^{-3}$ for the scaling model, and we use this as our final estimate. More generally, in the perturbative limit
($\delta_X, \delta_\alpha \ll 1$) the results of \cite{chluba1} imply $\delta_X^{\rm max} = -31 \delta_\alpha$ and therefore $\lambda' \simeq 1-29 \delta_\alpha$.

This increase in $X_e$ at fixed temperature compared to the canonical $\Lambda$CDM case can alternatively be thought of as slightly delaying the onset of photon decoupling and thus shifting the visibility function $g(z)$ to lower redshift. Very roughly, the shift in the centroid or peak redshift $z_p$ of $g(z)$ is of order $\Delta z_p \simeq -(2/3)(\delta_X+2\delta_\alpha)(1+z_p)$. As \cite{chluba1} show, the shape of the visibility function is left largely unchanged. For fixed $\Omega_b h^2$, the upward shift in the Hubble parameter, $H\rightarrow H(1+\delta_H)$, due to the dark sector in this model approximately restores $z_p$ and thus $g(z)$ to its canonical value if $\delta_X + 2\delta_\alpha \simeq \delta_H$ or $\delta_\alpha \simeq -0.034\delta_H$.


\subsection{Big Bang Nucleosynthesis}

In the mirror world model, the expansion rate at given temperature is $\lambda-1=8\%$ higher than in the Planck-$\Lambda$CDM model {\it at all temperatures}.  Thus, the weak interactions freeze out of equilibrium at higher temperature than in $\Lambda$CDM, leading to a higher primordial Helium abundance prediction from Big Bang Nucleosynthesis (BBN), $Y_p = 0.261 \pm 0.004$ \cite{cyrracine2021symmetry,2018CoPhC.233..237C}, than in the canonical model and in $3\sigma$ tension with the primordial Helium mass fraction inferred from observations, 
$Y_p = 0.2453 \pm 0.0034$ \cite{2020JCAP...03..010F,2021JCAP...03..027A}. Consistency with the observed Helium abundance could be restored by violating one or more additional assumptions of the standard cosmology. For example, through particle decay or annihilation, one could arrange entropy injection into the dark sector between the time of nucleosynthesis and photon decoupling (although this would violate the mirror symmetry between the two sectors) \cite{cyrracine2021symmetry,2022PhRvD.105l3516A}; in this way, one could have had $\lambda-1 \ll 0.08$ at the time of nucleosynthesis,  recovering the standard BBN prediction of $Y_p$, which is in good agreement with observation. Alternatively, as we did here for the electromagnetic fine-structure constant, one might invoke an appropriately boosted value of the weak coupling constant at the time of nucleosynthesis relative to the present; this would also lower the predicted BBN $Y_p$ value closer to that observed. Finally, we note that BBN places a weak upper bound on deviation of the fine-structure constant at redshift $z \sim 10^9$ from its current value, of order  $|\delta^{BBN}_\alpha| < 10^{-2}-10^{-1}$ \cite{2002PhRvD..66f3507N,2011LRR....14....2U}, which is consistent with the variation invoked here.

\section{\label{s:sec3}Scalar Field Model for time-varying $\alpha$}

We consider a simple model of an ultra-light scalar field $\phi$ phenomenologically coupled to electromagnetism, such that late-time classical evolution of the field is responsible for relaxation of $\alpha$ from its value at photon decoupling to its present value. We explore whether such a simple model is consistent with observed constraints.

\subsection{Scalar Field Coupling to Electromagnetism}

The Lagrangian for the scalar field is given by \cite{1982PhRvD..25.1527B,Olive_2002,Anchordoqui_2003}
\begin{equation}
    {\cal L}=\frac{1}{2}\partial_{\mu}\phi \partial^{\mu}\phi-V(\phi)-\frac{1}{4}Z_F\left(\frac{\phi}{M_{Pl}}\right)F_{\mu\nu}F^{\mu\nu} ~,
    \label{Lagrangian}
\end{equation}
where $Z_F$ is a dimensionless function of its argument, and the Planck mass $M_{Pl}=(8\pi G)^{-\frac{1}{2}}$. Following \cite{Anchordoqui_2003}, we have assumed that the scalar field does not couple appreciably to matter fields. 
Defining $\Delta\phi(t)=\phi(t)-\phi_0$, with $\phi_0 \equiv \phi(t_0)$ the present value of $\phi$, which we assume to be at or close to the minimum of its potential, and assuming that the deviation of $\phi(t)$ from its present value is small compared to the Planck mass at all times of interest, $\Delta \phi < M_{Pl}$, which ensures that quantum gravity corrections should be under 
control (see, e.g.,  \cite{Carroll_1998,Easther_2006}), we can expand the coupling term as 
\begin{equation}
    Z_F\left(\frac{\phi}{M_{Pl}}\right)=1+\kappa_1\frac{\Delta\phi}{M_{Pl}}+\kappa_2\left(\frac{\Delta\phi}{M_{Pl}}\right)^2+... ~.
    \label{ZF}
\end{equation}
Assuming $\delta_\alpha \ll 1$, from the RHS of Eqn.(\ref{ZF}) we have \cite{Olive_2002} 
\begin{equation}
\delta_\alpha =\frac{\Delta\alpha}{\alpha} \simeq -\kappa_1\frac{\Delta\phi}{M_{Pl}}-\left(\kappa_2-\kappa_1^2\right)\left(\frac{\Delta\phi}{M_{Pl}}\right)^2 
\label{dadp}
\end{equation}
to quadratic order in $\Delta \phi/M_{Pl}$.  

We consider classical evolution of the scalar field in the expanding universe, assuming it to be approximately homogeneous over scales of interest, in which case 
\begin{equation}
\ddot\phi+3H\dot\phi=-\frac{\partial
V}{\partial\phi}~,
\end{equation}
and the energy density of the field is given by 
\begin{equation}
\rho_\phi=\frac{1}{2}\dot\phi^2+V(\phi)~.
\end{equation}
While one could consider a variety of models for the scalar field potential $V(\phi)$, here we focus only on the simplest 
case of a free, massive field, with 
$V(\phi)=m^2\phi^2/2$. We assume that self-interactions or quantum corrections to $V(\phi)$ do not generate terms large compared to the mass term; this requires an extremely small upper bound on the quartic self-coupling of the field, as is the case with axion-like fields. We make no attempt here to embed $\phi$ into a fundamental theory. 

With these assumptions, the scalar equation of motion becomes 
\begin{equation}
\ddot\phi+3H\dot\phi+m\phi^2=0~.
\label{EOM}
\end{equation}
   We now consider various phases in and constraints upon the evolution of $\phi$, subject to the constraint that it implement the scaling solution for $\lambda'$ before and during photon decoupling.

We focus on the simple model of $\alpha$-evolution discussed in \S \ref{s:sec2}, in which $\delta_ \alpha=-2.7\times 10^{-3}$ is a non-zero constant prior to the time of photon decoupling and subsequently relaxes to zero. From Eqn.(\ref{dadp}), this implies $\Delta \phi=$ constant until at least $z_{\rm dec}=1100$: the scalar field must be frozen until the time that photon decoupling is nearly completed. From Eqn.(\ref{EOM}), at early times, when the expansion rate $H \gg m$, the solution is indeed $\phi=$ constant. The field begins to evolve when $H \simeq 2m/3$, which gives an approximate upper bound on the scalar mass 
\begin{equation}
    m < \frac{3}{2}H(T_{\rm dec}) \simeq \frac{3}{2}H_0 \Omega_m^{0.5} (1+z_{\rm dec})^{3/2} \simeq 4\times 10^{-29} ~{\rm eV}~.
    \label{m}
\end{equation}

From Eqn.(\ref{m}), the field starts to evolve at redshift $z_c$ given by 
\begin{equation}
    1+z_c \simeq 1100 \left(\frac{m}{4\times 10^{-29} ~{\rm eV}}\right)^{2/3}~.
    \label{zc}
\end{equation}
At redshifts $z<z_c$, the field undergoes damped oscillations around $\phi=0$, and its oscillation-average energy density redshifts like non-relativistic matter,
\begin{eqnarray}
\rho_\phi(z<z_c) \simeq && \frac{1}{2}m^2 \phi_i^2\left(\frac{1+z}{1+z_c}\right)^{3} \nonumber \\
= && 0.005\left(\frac{\phi_i}{M_{Pl}}\right)^2 \left(\frac{1+z}{1100}\right)^3 ~{\rm eV}^4~,
\label{rhophi}
\end{eqnarray}
where $\phi_i \simeq \phi(t_c)$ is the initial field amplitude. 

An important constraint is that the energy density of $\phi$ must be at all times subdominant compared to that of other species that contribute significantly to the energy density of the Universe: in this model, $\phi$ is {\it not} the dark energy component. Its energy density relative to matter reaches a maximum at $z_c$ (and is constant thereafter), at which point $\rho_\phi(z_c) \simeq \rho(\phi_i) = m^2 \phi_i^2/2$. Since the matter density at that time is given by $\rho_m = \Omega_m \rho_{\rm crit}(1+z_c)^3$, using Eqn.(\ref{zc}) gives
\begin{equation}
   \frac{\rho_\phi(z_c)}{\rho_m(z_c)} \simeq 0.3\left(\frac{\phi_i}{M_{Pl}}\right)^2 ~,
   \label{denratio}
\end{equation}
independent of $m$ or $z_c$. Thus the energy density in the scalar field remains subdominant compared to matter provided that the field excursion is sufficiently sub-Planckian. For example, requiring that the energy density in the scalar field be smaller than that in the mirror dark sector implies $|\phi_i|/M_{Pl}< 0.5$, a relatively mild constraint.

The scalar field oscillation-average amplitude decays as 
\begin{equation}
    \phi(z) \simeq \phi_i \left(\frac{1+z}{1+z_c}\right)^{3/2}~.
    \label{phiz}
    \end{equation}
Thus, from Eqn.(\ref{dadp}), assuming $\phi_0 \ll \phi_i$,
we have the initial condition
\begin{equation}
2.7\times 10^{-3} \simeq \kappa_1\frac{\phi_i}{M_{Pl}}+\left(\kappa_2-\kappa_1^2\right)\left(\frac{\phi_i}{M_{Pl}}\right)^2 ~,
\label{dadp2}
\end{equation}
and at later times, $z<z_c$, the fractional deviation of $\alpha$ from its present value is given by 
\begin{eqnarray}
    \delta_\alpha(z)= && ~ \kappa_1\frac{\phi_i}{M_{Pl}}\left(\frac{1+z}{1+z_c}\right)^{3/2} \nonumber \\
    +&& \left(\kappa_2-\kappa_1^2\right)\left(\frac{\phi_i}{M_{Pl}}\right)^2 \left(\frac{1+z}{1+z_c}\right)^{3} ~.
    \label{dalphaz}
\end{eqnarray}

We consider two qualitatively different parameter regimes for the scalar coupling to electromagnetism: (1) $\kappa_1 \gg \kappa_2(\phi_i/M_{Pl})$, i.e., the term linear in $\phi$ dominates in Eqn.(\ref{dalphaz}) for all times, and (2) $\kappa_1 \ll \kappa_2(\phi_i/M_{Pl})(1+z_c)^{-3/2}$, in which case the quadratic term always dominates. For case (1), from Eqn.(\ref{dadp2}), $\kappa_1(\phi_i/M_{Pl})=2.7\times 10^{-3}$, which from the field excursion limit above implies the lower bound $|\kappa_1|> 5.4\times 10^{-3}$. For case (2), we instead have $\kappa_2 (\phi_i/M_{Pl})^2=2.7\times 10^{-3}$, with the lower bound $\kappa_2 > 0.01$ from Eqn.(\ref{denratio}).

\subsection{Scalar Field Constraints from Observational Bounds on Time-variation of $\alpha$}

Observations at late times impose strict bounds on $\Delta \alpha/\alpha$ at low redshift. We separately consider bounds on cases (1) and (2) as defined above, i.e., for linear and quadratic scalar coupling to electromagnetism.

\subsubsection{Case (1): linear coupling $\kappa_1$}

Observational constraints can be couched in terms of an upper bound on the fractional deviation in $\alpha$ at redshift $z_x$, which we denote by $\delta^{\rm max}_\alpha(z_x)$.
In the case that the linear $\kappa_1$ term dominates, then from Eqns.(\ref{phiz}, \ref{dalphaz}), a late-time constraint at redshift $z_x$ translates to 
\begin{equation}
    1+z_c > \left(\left|\frac{\delta_{\alpha}(z_{\rm dec})}{\delta^{\rm max}_{\alpha}(z_x)}\right|\right)^{2/3} (1+z_{\rm x})~,
    \label{eqzc}
\end{equation}
where from \S 2 we have $\delta_\alpha(z_{\rm dec})=-2.7\times 10^{-3}$.
For example, consistency with QSO spectra yields $\delta^{\rm max}_\alpha \simeq 10^{-5}$ at redshift $z_x \simeq 4$ \cite{2011LRR....14....2U}. 
From Eqn.(\ref{eqzc}) this implies $z_c > 250$; from Eqn.(\ref{zc}), this yields a lower bound on the scalar field mass,
\begin{equation}
    m>4 \times 10^{-30} ~{\rm eV} ~~~{\rm (QSO)}.
    \label{mmin}
\end{equation}
Similarly, the meteorite bound, $\delta^{\rm max}_\alpha(z=0.45) =3\times 10^{-7}$ \cite{Anchordoqui_2003}, implies $1+z_c > 627$, or 
\begin{equation}
 m>1.7 \times 10^{-29} ~{\rm eV} ~~~{\rm (meteorite)}.
 \label{mmin2}
\end{equation}
Finally, the bound from element ratios in the Oklo natural reator, $\delta^{\rm max}_\alpha=10^{-7}$ at $z=0.14$, gives $1+z_c>1025$ and 
\begin{equation}
    m> 3.6\times 10^{-29} ~{\rm eV}~~~~{\rm (Oklo)}~.
    \label{mmin3}
\end{equation}
From Eqns.(\ref{m}-\ref{mmin3}) we see that the scalar field mass is constrained to the narrow range $m=(3.6-4) \times 10^{-29}$ eV, and the redshift $z_c$ when the field begins to oscillate is constrained to $1025 < 1+z_c < 1100$, with the tightest late-time constraints coming from the Oklo natural reactor.

\subsubsection{Case (2): quadratic coupling $\kappa_2$}

The scalar mass constraints derived above assume that the linear $\kappa_1$ term dominates in Eqn. (\ref{ZF}). If an approximate symmetry forbids the linear term or makes it subdominant, then the quadratic term drives a more rapid transition in $\alpha$ for given scalar field evolution $\phi(t)$ and yields consequently weaker constraints on $m$ and $z_c$ from observational bounds on $\alpha$ variation. Considering only the quadratic term in Eqn.(\ref{ZF}), the late-time constraints become
\begin{equation}
    1+z_c > \left(\left|\frac{\delta_{\alpha}(z_{\rm dec})}{\delta^{\rm max}_{\alpha}(z_x)}\right|\right)^{1/3} (1+z_{\rm x})~,
    \label{eq:zminquad}
\end{equation}
for constraints at $z=z_x$. The corresponding $z_c$ and $m$ constraints become: $z_c>32$ and $m>2\times 10^{-31}$ eV (QSOs); $z_c>30$ and $m>1.8\times 10^{-31}$ eV (meteorites); $z_c>34$ and $m>2.2\times 10^{-31}$ eV (Oklo). In this case, all three constraints  provide comparable bounds on the scalar mass, which  has an allowed range of approximately two orders of magnitude.

 
 \subsubsection{Constraints on $V(\phi)$}

Finally, we note that other choices for the form of the scalar field potential, $V(\phi)$, lead to different evolutionary behaviors once $H<m$, which would change the constraints above. For example, for a monomial potential of the form $V(\phi) \propto \phi^{2n}$, the oscillation-average equation of state parameter for the field is given by $\omega_\phi\simeq\ (n-1)/(n+1)$, the average energy density of the oscillating field redshifts as $\rho_\phi\propto a^{-3(1+\omega_\phi)}$ \cite{PhysRevD.28.1243}, which is faster than that of non-relativistic matter for $n>1$, but the field amplitude redshifts as $\phi =\phi_i [(1+z)/(1+z_c)]^{3/(n+1)}$, which is slower than for the $n=1$ case discussed above. Given the very narrow range of allowed scalar mass in the $n=1$ case above for linear coupling to electromagnetism, values of $n>1$ appear to be excluded by the late-time constraints on $\Delta \alpha$ unless the electromagnetic coupling is quadratic or higher.

\subsection{Constraints from Equivalence Principle Tests}

Following earlier work of Dicke, Beckenstein \cite{1982PhRvD..25.1527B} noted that, if the electromagnetic coupling involves a dynamical field that can vary in space as well as time, then it can lead to composition-dependent inertial forces that violate the very precise tests of the Equivalence Principle. Olive and Pospelov \cite{Olive_2002} used this to derive constraints on the linear and quadratic coupling terms in Eqn.(\ref{ZF}). Specifically, they derived an upper bound 
that corresponds to $|\kappa_1|< 10^{-3}$ from differential acceleration of the Earth and the Moon toward the Sun. This is in conflict with the lower bound of $\kappa_1>5.4\times 10^{-3}$ derived from the scaling solution to the Hubble tension and the energy density of the scalar field. Moreover, for values of $\phi_i \ll M_{Pl}$, this lower bound on $\kappa_1$ is correspondingly larger. As a consequence, the linear coupling model appears to be disfavored, at least in the context of a single massive, free scalar field.

For case (2), Olive and Pospelov \cite{Olive_2002} find that the differential acceleration is approximately (translated to our notation) 
\begin{equation}
    \frac{\Delta g}{g} \simeq 8\times 10^{-6} \kappa_2^2 \left(\frac{\phi_0}{M_{Pl}}\right)^2 f~,
\end{equation}
where $f$ is an order unity function that depends on the composition difference between two test masses in a gravitational field. The MICROSCOPE collaboration recently obtained a stringent constraint on the differential acceleration of titanium and platinum in orbit around the Earth, $\Delta g/g < 10^{-15}$, over an order of magnitude stronger than previous constraints \cite{PhysRevLett.129.121102}. Imposing this bound and using Eqn.(\ref{phiz}), we find the constraint
\begin{equation}
    1+z_c > 2000 f^{1/3} \left(\kappa_2 \frac{\phi_i}{M_{Pl}}\right)^{2/3} \simeq 278(\kappa_2 f)^{1/3}~,
\end{equation}
where the second equality comes from the initial condition for $\delta_\alpha$. Since the energy density bound is $\kappa_2>0.01$, and to reasonable approximation we can take $f^{1/3} \simeq 1$, this provides the constraint $z_c>60$ in this model. This lower bound on $z_c$, which corresponds to a scalar mass bound $m>5\times 10^{-31}$ eV, is stronger than the bounds from $\alpha$ variation at late times for this case ($z_c>34$ from Eqn. (\ref{eq:zminquad})). However, even if we were to require $\phi_i \ll M_{Pl}$ and thus larger values of $\kappa_2$ for the Hubble tension solution, this bound does not close off a very large portion of the parameter space, since the scaling of the lower bound on $z_c$ with $\kappa_2$ is weak.

\section{Conclusion}
We have developed a variant of the mirror world dark sector model proposed in \cite{cyrracine2021symmetry} to resolve the Hubble tension between CMB and large-scale structure measurements in the context of $\Lambda$CDM and local measurements of the expansion rate. We replace the ad hoc adjustment of the primordial Helium abundance in \cite{cyrracine2021symmetry}, which disagrees with both observation and the self-consistent prediction of Big Bang Nucleosynthesis, with a dynamical model for evolution of the electromagnetic fine-structure constant, $\alpha$, prior to photon decoupling. The scaling symmetry exploited in \cite{cyrracine2021symmetry} to resolve the Hubble tension is approximately realized if the value of $\alpha$ prior to photon decoupling is lower than its current value by $\Delta \alpha = -2 \times 10^{-5}$. This shift boosts the photon inverse mean free path primarily by reducing the binding energy of atomic Hydrogen at early times by 0.07 eV, which increases the free electron number density $n_e$ at fixed temperature. The requisite change in $\alpha$ is smaller than one might naively expect, due to the strong sensitivity of the free electron density to the H binding energy.

We then considered a simple scalar field model with phenomenological coupling to electromagnetism as a toy model that instantiates $\alpha$-evolution after photon decoupling via classical relaxation of the field. For the case of linear field coupling, consistency with stringent late-time constraints on variation of $\alpha$ from its current value constrains the mass of the field to the relatively narrow range $m=(3.6-4)\times 10^{-29}$ eV, corresponding to a critical redshift range $z_c=1025-1100$. However, in this case, Equivalence Principle tests place an upper bound on the value of the linear coupling constant $\kappa_1<10^{-3}$ that conflicts with the lower bound required by the Hubble tension solution and the scalar field energy density, $\kappa_1> 5.4\times 10^{-3}$, so this version of the model appears to be disfavored. For quadratic coupling of the field to electromagnetism, late-time $\alpha$-variation constraints bound the scalar field mass to the range $2.6\times 10^{-31} ~{\rm eV} < m=4\times 10^{-29}$ eV, which implies that the field and $\alpha$ begin relaxing toward their current values in the redshift range $37<z_c \sim 1100$. In this case, the recent Equivalence Principle limit, in combination with the requirement that the scalar field energy density be subdominant compared to that of the dark sector, provides a stronger constraint, $5\times 10^{-31} ~{\rm eV} < m=4\times 10^{-29}$ eV and  $60<z_c \sim 1100$.

A feature of this model, as in that of \cite{cyrracine2021symmetry}, is that the physical origin of the $\lambda'$ shift in the photon scattering rate (dynamical evolution of $\alpha$) is completely different from that of the $\lambda$ shift in density and thus $H_0$ (the mirror dark sector), so there is no symmetry or physical mechanism that drives the requirement $\lambda'=\lambda$. One hopes that a more compelling version of the model could be found that provides a rationale for this coincidence.

\section{Acknowledgements}

We thank Lloyd Knox for helpful discussion, and we thank the referee for helpful comments that have improved the paper. This work was supported by the Department of Energy grant DE-AC02-07CH11359 subcontract 6749003 at Fermilab and the University of Chicago. JF also acknowledges the hospitality of the Aspen Center for Physics, supported by NSF grant PHY-1607611, where part of this work was carried out.

\bibliography{paper}
	
\end{document}